\documentclass[a4paper]{article}
\usepackage[margin=25mm]{geometry}
\usepackage[utf8]{inputenc}
\usepackage{authblk}
\usepackage[shortlabels]{enumitem}
\usepackage{comment}

\usepackage[shortlabels]{enumitem}
\usepackage{hyperref}
\def\ind{\mathbbm{1}}

\usepackage{%
	amsfonts,%
	amssymb,%
	amsthm,%
	bbm,%
	float,%
	etoolbox,%
	mathrsfs,%
	mathtools,%
	multirow,%
	pgf,%
	subfigure,%
	tikz,%
        stfloats,
        booktabs,
}

\allowdisplaybreaks  

\usepackage{multirow,tabularx}
\newcolumntype{Y}{>{\centering\arraybackslash}X}
\renewcommand\arraystretch{2}
\newcolumntype{?}{!{\vrule width 1pt}}

\theoremstyle{plain}
\newtheorem{thm}{Theorem}
\newtheorem{lem}{Lemma}
\newtheorem{proposition}{Proposition}

\newtheorem{remark}{Remark}

\theoremstyle{definition}

\newtheoremstyle{case}{}{}{}{}{}{:}{ }{}
\theoremstyle{case}

\theoremstyle{remark}

\newcommand{\norm}[1]{\left\lVert#1\right\rVert}

\newcommand{\E}{\mathbb{E}}

\newcommand{\mb}[1]{\mathbb{#1}}

\newcommand{\brac}[1]{\left(#1\right)}

\newcommand{\sbrac}[1]{\left[#1\right]}

\newcommand{\expect}[1]{\mathbb{E}\sbrac{#1}}

\newcommand{\abs}[1]{\lvert #1 \rvert}

\providecommand{\keywords}[1]
{
  \small	
  \textbf{\textit{Keywords---}} #1
}

\title{On Optimal Server Allocation for Moldable Jobs with Concave Speed-Up}

\author[1]{Samira Ghanbarian}
\author[2]{Arpan Mukhopadhyay}
\author[3]{Ravi R. Mazumdar}
\author[4]{Fabrice M. Guillemin}

\affil[1,3]{University of Waterloo, Waterloo, Canada}
\affil[2]{University of Warwick, Coventry, U.K.}
\affil[3]{Orange Innovation, Lannion, France}

\date{} 

\begin{document}
\maketitle

\begin{abstract}
A large proportion of jobs submitted to modern computing clusters and data centers are parallelizable and capable of running on a flexible number of computing cores or servers. 
Although allocating more servers to such a job results in a higher speed-up in the job's execution, it reduces the number of servers available to other jobs, which in the worst case, can result in an incoming job not finding any available server to run immediately upon arrival.
Hence, a key question to address is: how to optimally allocate servers to jobs such that (i) the average execution time across jobs is minimized and (ii) almost all jobs find at least one server immediately upon arrival. 
To address this question, we consider a system with $n$ servers, where jobs are parallelizable up to $d^{(n)}$ servers and the speed-up function of jobs is concave and increasing. Jobs not finding any available servers upon entry are blocked and lost.
We propose a simple server allocation scheme that achieves the minimum average execution time of accepted jobs while ensuring that the blocking probability of jobs vanishes as the system becomes large ($n \to \infty$). This result is established for various traffic conditions as well as for heterogeneous workloads.
To prove our result, we employ Stein's method which also yields non-asymptotic bounds on the blocking probability and the mean execution time. Furthermore, our simulations show that the performance of the scheme is insensitive to the distribution of job execution times.
\end{abstract}

\maketitle

\section{Introduction}
The demand for large-scale computations on cloud-based platforms has seen a steady growth in the past decade~\cite{quasar}.
Many high performance computing  (HPC) jobs including training of machine learning models~\cite{TensorFlow_2016}, simulation of climate models~\cite{climate}, prediction of protein structures using protein folding~\cite{protein}, and processing large database queries~\cite{database}
are now performed on cloud resources. 
Most of these HPC jobs are highly parallelizable and hence can run on a large number of cores or servers providing parallelization gains to the jobs' execution. Empirical studies show that the majority of these parallelizable jobs are {\em moldable}~\cite{moldable3}. A moldable job is a parallelizable job that can run on a flexible number of servers and, therefore, its execution time is determined by the number of servers allocated to it upon arrival. Thus, unlike {\em rigid} jobs, moldable jobs can easily adapt to the number of available servers in the system and therefore, have the potential to significantly improve the system's overall resource utilization~\cite{moldable1,moldable2}.


A moldable job is characterized by its {\em speed-up function} which expresses the reduction in the job's execution time as a function of the number of servers allocated to it. The speed-up function is typically a concave and strictly increasing function of the number of allocated servers. 
This observation stems from Amdahl's law \cite{Amdahl's_law_2008}, which states that if a fraction $p\in [0,1]$ of a job is parallelizable, then running it on $i$ servers yields a speed-up of $\frac{1}{(1-p)+p/i}$, a concave and increasing function of $i$. 
While allocating more servers to moldable jobs results in lowering their execution times, it reduces the number of servers available to future jobs, thereby increasing the chance of a future job not finding any available server on arrival. In many modern cloud-based cluster management systems, jobs that cannot find enough resources immediately upon arrival are {\em blocked}~\cite{Google_Borg_2015,autopilot_google}, resulting in a degradation in the quality of users' experience. 
Therefore, the key question we address in this paper is:
{\em How to optimally allocate servers to jobs such that the average execution time of jobs is minimized while ensuring that no job is blocked?}

To address this question, we consider a loss system with $n$ unit-rate servers, where the submitted jobs are parallelizable up to $d^{(n)}$ servers. The threshold $d^{(n)}$ captures the maximum degree to which a job can be parallelized. 
We assume that the threshold $d^{(n)}$ can potentially be large and can scale with the system size $n$ to allow for massively parallelizable jobs~\cite{mpc1,mpc2}. The speed-up function of a job is represented by a vector $s=(s_i, i \in \{0,1,\ldots, d^{(n)}\})$ where for each $i\in \{0,1,\ldots, d^{(n)}\}$, $s_i$ denotes the factor by which the job's execution time is reduced if it is run on $i$ servers compared to its execution time on a single server. Following Amdahl's law, we assume that the speed-up function is concave and strictly increasing with respect to the number of allocated servers. 
Jobs that cannot find any available (idle) server immediately upon arrival are assumed to be \textit{blocked}. This is in line with the admission control policies employed in large-scale cluster management frameworks such as Google Borg~\cite{Google_Borg_2015}.
Our objective is to design server allocation schemes that (i) minimize the average execution time of accepted jobs and (ii) ensure that the blocking probability remains `close' to zero (i.e., almost all jobs are accepted into the system). Henceforth, we shall refer to these two criteria together as the {\em optimality criteria}. Toward achieving these criteria, we make the following contributions:

{\bf (1)} We first find a sufficient condition for any {\em online} server allocation scheme to satisfy the optimality criteria. 
The sufficient condition states that if the steady-state values under a scheme solve a specific convex optimization problem, then the scheme satisfies the optimality criteria described above. 

{\bf (2)} Designing schemes which achieve a target steady-state behavior is not easy (such schemes may not even exist in general). Thus, to facilitate the construction of desired schemes, we investigate the structural properties of the optimal solution. 
Using the properties of the speed-up function, we show that the optimal solution exhibits a {\em state space collapse} (SSC) property wherein the dimension of the optimal solution reduces from $d^{(n)}$ to at most {\em two}. 


{\bf (3)} Based on the above SSC property, we introduce a probabilistic greedy server 
allocation scheme 
and show that the scheme achieves the desired optimality criteria asymptotically as the system size $n$ becomes large - a property which we refer to as {\em asymptotic optimality}. We establish this property under both light and heavy traffic regimes and heterogeneous workloads. 
As a corollary of our result, it follows that when the speed-up function of jobs is perfectly linear, a greedy scheme which always allocates the maximum possible number of available servers to each incoming job is asymptotically optimal. However, this scheme is no longer optimal when the speed-up is sub-linear. In the latter case,  we show how the optimal number of servers to allocate to each job depends on the arrival rate of jobs and the speed-up function. 
Furthermore, through numerical simulations, we demonstrate that the proposed scheme is nearly insensitive to job size distributions for large system sizes. 

{\bf (4)} To establish our theoretical results, we employ Stein's method to compare the generator of the underlying Markov process to that of a simpler dynamical system. 
In addition to proving asymptotic optimality, this approach provides non-asymptotic bounds that characterize the rate at which convergence takes place.
This approach is similar to the approach found in recent works~\cite{lei_ying_2020_Stein,Srikant_Stein_2020}. 
However, one significant difference from earlier works is that in our model the jobs run on multiple servers simultaneously as opposed to the single-server jobs considered in earlier works. This makes finding the appropriate Lyapunov functions for demonstrating state space collapse and state-space concentration much more challenging. We believe that our approach is useful for analyzing other similar models of multi-server jobs. 
{\bf Related work}: Motivated by the prevalence of complex jobs in data centers, there has recently been an interest in the scheduling of {\em multi-server} jobs, broadly defined as jobs that can run on multiple servers simultaneously (instead of only a single server).
Prior research on multi-server jobs has mainly focused on jobs requiring a fixed number of servers - also referred to as {\em rigid jobs}. In \cite{Brill_multiserver_1984} and \cite{FILIPPOPOULOs_multiserver_2007}, the stationary distribution of job numbers in a system with 
two servers under the FCFS policy was studied. Extending to systems with an arbitrary number of servers, \cite{multiserver_asymptotic_opimality_2022} introduced the ServerFilling-SRPT policy, 
minimizing the mean response time under heavy traffic limits. 
Additionally, \cite{GROSOF_Reset} addressed the mean response time under FCFS policy, characterizing it up to an additive constant, while works such as \cite{Zero_wait_central-2021} and \cite{sharp_zer_central_2022} studied asymptotic regimes where the number of servers tends to infinity, showing a zero asymptotic waiting time.

Another line of work, more closely related to the present paper, studies {\em flexible} multi-server jobs which do not have fixed server requirements; instead, such jobs can run on a flexible number of servers with execution times characterized by a speed-up function. 
In \cite{first_speed_up}, 
a scheduling scheme called EQUI was introduced for flexible jobs. In the EQUI scheme, at any point of time, all servers are evenly distributed among existing jobs. Although EQUI has a good competitive ratio, it is {\em preemptive} in that it requires the allocation of servers to change within the lifetime of a job.
To address this issue, 
\cite{Berg_speedup_2018} proposed a fixed width policy, which {\em approximately} matches the mean response time of the EQUI policy in the stochastic setting. Unlike their model, which considers processor sharing (PS) servers with no losses, in our model a server cannot be shared simultaneously by multiple jobs and jobs may be lost if all servers are found busy. Furthermore, our analysis characterizes 
the exact allocation of servers as a function of the speed-up function of jobs and the arrival rate as opposed to their analysis which only provides an approximate solution. Other related works such as \cite{Berg_speedup_2020,Berge_different_phase_2021,WCFS_Grososf_2022} consider similar multi-server job models. However, these works consider queue-based scheduling of jobs with perfectly linear speed-up functions as opposed to the current setting where there is no queue and the speed-up can be any concave and increasing function of the number of allocated servers. 

The model closest to ours is the classical multi-rate Erlang loss model~\cite{kelly_loss,mazum_blocking}. However, in this model, jobs have fixed server demands independent of the state of the system seen at arrival instants. Hence, this model differs from our model where the the number of allocated servers as well as the completion time of a job becomes {\em state-dependent}. The dependency on the state makes our analysis significantly more challenging.

\textbf{Organization.} The rest of this paper is organized as follows. In Section~\ref{section: system model} we introduce the system model. In Section ~\ref{sec:lower}, we derive a sufficient condition for asymptotic optimality. The proposed scheme is introduced and analyzed in Sections~\ref{sec: main results} and~\ref{section: proofs}, respectively. Section~\ref{sec:het} discusses the generalization to heterogeneous workloads. Numerical results are provided in Section~\ref{section: numerical results}.

\section{System model} 
\label{section: system model}

Consider a system with $n$ servers where jobs arrive according to a Poisson process with rate $n\lambda^{(n)}$. We assume $\lambda^{(n)}$ varies with the system size $n$ as $\lambda^{(n)}=1-\beta n^{-\alpha}$, where $\alpha,\beta >0$ are positive constants such that $\lambda^{(n)} \in (0,1)$. This allows us to study system performance under different traffic conditions: (i) the mean-field regime which corresponds to
$\alpha = 0, \beta \in (0, 1)$, (ii) the Halfin-Whitt regime which corresponds to $\alpha = 1/2, \beta >0$, (iii) the sub-Halfin-Whitt (resp. super-Halfin-Whitt) regime corresponding to
$\alpha \in (0, 1/2), \beta>0$ (resp. $\alpha \in (1/2, 1), \beta>0$), and (iv) the super-non-degenerate slowdown (NDS) regime corresponding to $\alpha \geq 1$. 

Each job can run on any number of servers in the range $[1,d^{(n)}]$, with each server capable of processing at most one job at any given time. If a job is allocated $i \in [1,d^{(n)}]$ servers, its execution time decreases by a factor of $s_i$ compared to running on a single server, with $s_0=0$ for completeness. We assume that the speed-up function of jobs, denoted by $s=(s_i, i\in \{0,1,\ldots,d^{(n)}\})$, satisfies the following properties:
\begin{enumerate}[{\bf P1}]
    \item \label{prop:increasing}The speed-up function is strictly increasing and satisfies
    \begin{equation}
        0=s_0 < 1= s_1 < s_2 < \ldots < s_{d^{(n)}}.
        \label{eq:increasing}
    \end{equation}
    \item \label{prop:concavity}The speed-up function is concave and, therefore, satisfies
    \begin{equation}
        1=\frac{s_1}{1} \geq \frac{s_2}{2} \geq \ldots \geq \frac{s_{d^{(n)}}}{d^{(n)}}.
    \label{eq:concave}
    \end{equation}
\end{enumerate}
%
We will distinguish between the {\em linear speed-up} case for which $s_i=i$ for all $i \in [1,d^{(n)}]$, and the {\em sub-linear speed-up} case for which there exists $i \in [1,d^{(n)}]$ such that $s_i < i$.

We assume that a job's inherent execution time (i.e., the execution time on a single server) is an exponential random variable with unit mean, independent of other jobs' execution times and the arrival process. We will investigate the effect of different execution time distributions later in the paper.

Upon a job arrival, if no server is found available, the job is {\em blocked} which corresponds to a {\em loss}; otherwise, if at least one server is found available, a {\em server allocation scheme} is used to determine the number of available servers to be allocated to the job. The allocated servers remain {\em occupied} as long as the job executes. 

Allocating more available servers to jobs decreases their average execution time, but it also increases the probability of future arrivals being blocked due to fewer available servers. On the other hand, allocating fewer available servers to jobs reduces the blocking probability at the expense of higher average delay. In this paper, we aim to design allocation schemes which achieve two objectives simultaneously: (i) minimize the average execution time of jobs and, (ii) maintain a near-zero blocking probability\footnote{Achieving exactly zero blocking may be infeasible for a finite system with stochastic arrivals and service times as under any allocation scheme there will be non-zero probability with which all servers become busy.}. Specifically, we aim to design schemes achieving the {\em minimum average execution time} for accepted jobs while ensuring that the {\em blocking probability} approaches zero as system size increases ($n \to \infty$). We shall refer to such schemes as {\em asymptotically optimal} schemes as no scheme which achieves zero blocking (asymptotically) can achieve a smaller average execution time.

{\bf Notations and State Descriptor}: Throughout our analysis, we use the following notations. For any integer $d$, we denote by $[d]$ the set $\{1,2\ldots,d\}$, and we denote by $\norm{\cdot}$, the $\ell_1$-norm on $\mathbb R^d$. For each $i \in [d^{(n)}]$, we let $X_i^{(n)}(t)$ denote the number of jobs running on $i$ servers simultaneously at time $t\geq 0$. We define $x_i^{(n)}(t), i\in [d^{(n)}],$ as $x_i^{(n)}(t)=X_i^{(n)}(t)/n$. Clearly, under any online allocation scheme, the process $x^{(n)}(\cdot)=(x_i^{(n)}(\cdot), i\in [d^{(n)}])$ has a unique stationary distribution. By omitting the explicit dependence on $t$, we let $x^{(n)}=(x_i^{(n)}, i\in [d^{(n)}])$ denote the state of the system distributed according to its stationary distribution. Additionally, we define the fraction of busy servers at steady-state as $q_1^{(n)} := q_1(x^{(n)}) = \sum_{i \in [d^{(n)}]} ix_i^{(n)} \in [0,1]$, and the fraction of idle servers at steady-state as $q_0^{(n)}:=q_0(x^{(n)}) = 1 - q_1(x^{(n)})$. We also define  $r^{(n)}:=r(x^{(n)})=\sum_{i \in [d^{(n)}]} s_i x_i^{(n)}$ to be the (normalized) rate of departure of jobs from the system in the steady-state. We note that $r^{(n)}=q_1^{(n)}$ when the speed-up function of jobs is linear. We let $P_b^{(n)}$ denote the steady-state blocking probability of jobs and $D^{(n)}$ denote the random variable having the same distribution as the execution time of a job in the steady-state. We will omit the superscript $\vphantom{\cdot}^{(n)}$ from our notations when from the context it is clear that we are considering a finite system of size $n$.

\section{Optimality criterion}
\label{sec:lower}

For any server allocation scheme, the steady-state quantities must satisfy the following equations.
\begin{align}
    &\lambda^{(n)}\brac{1-P_b^{(n)}}=\expect{r^{(n)}}=\sum_{i\in [d^{(n)}]}s_i \expect{x_i^{(n)}},\label{eq:rate_consevation}\\
    &\lambda^{(n)}\brac{1-P_b^{(n)}}\expect{D^{(n)}}=\sum_{i\in [d^{(n)}]} \expect{x_i^{(n)}}\label{eq:little}.
\end{align}
The first equation follows from the rate conservation principle at steady-state, 
and the second equation follows from Little's law. 
Thus, for a server allocation scheme to achieve zero blocking (i.e., $P_b^{(n)}=0$) we must have
\begin{align}    &\lambda^{(n)}=\expect{r^{(n)}}=\sum_{i\in [d^{(n)}]}s_i \expect{x_i^{(n)}}\label{eq:rate_consevation_zerob},\\  &\lambda^{(n)}\expect{D^{(n)}}=\sum_{i\in [d^{(n)}]} \expect{x_i^{(n)}}\label{eq:little_zerob}.
\end{align}
Note that for such a scheme, the average execution time $\mathbb{E}[D^{(n)}]$ is proportional to the (scaled) expected number of jobs $\sum_{i \in [d^{(n)}]}\mathbb{E}[x_i^{(n)}]$ in the system.
Hence, to further minimize the average execution time at steady-state, the steady-state expectations $\mathbb{E}[x_i^{(n)}], i\in [d^{(n)}]$, must be a solution of the following linear program in $y=(y_i, i\in [d^{(n)}])$:
\begin{equation}
        \begin{aligned}
        & \underset{y=(y_i,i\in [d^{(n)}])} {\text{minimize}}
        & & \frac{1}{\lambda^{(n)}}\sum_{i \in [d^{(n)}]} y_i\\
        & \text{subject to}
        & & r(y)=\sum_{i \in [d^{(n)}]} s_i y_i = \lambda^{(n)}, \\
        & & &q_1(y)=\sum_{i\in [d^{(n)}]} i y_i \leq 1,\\
        & & &  y_i \geq 0, \quad \forall i \in [d^{(n)}],
        \end{aligned}
        \tag{P}
    \label{eq:opt}
\end{equation}
where $\E[x_i^{(n)}]$ is replaced with $y_i$ for each $i \in [d^{(n)}]$. The first inequality constraint in~\eqref{eq:opt} follows from the fact that $\E[q_1^{(n)}]= \sum_{i\in [d^{(n)}]}i \E[x_i^{(n)}] \leq 1$.
Let $D^{*,(n)}$ denote the optimal objective function value of~\eqref{eq:opt}. We obtain a sufficient condition for asymptotic optimality as stated in the following proposition whose formal poof is provided in the appendix.

\begin{proposition}  
\label{proposition: optimality}
Problem~\eqref{eq:opt} has at least one optimal solution if and only if $\lambda^{(n)} \leq 1$. 
%
Let $\lambda^{(n)} \in (0,1]$ for all $n \in \mathbb{N}$. Then, a server allocation scheme is asymptotically (as $n \to \infty$) optimal, i.e., it achieves zero blocking (i.e., $P_b^{(n)} \to 0$) and minimum average execution time of jobs (i.e., $\abs{\E[D^{(n)}]-D^{*,(n)}} \to 0$) in the steady-state if under that scheme the steady-state expectations satisfy $\mathbb{E}[\vert \vert x^{(n)}-y^{*,(n)}\vert\vert] \to 0$ as $n \to \infty$, where $y^{*,(n)}$ denotes an optimal solution of~\eqref{eq:opt}. 
\end{proposition}
%
%
%

Proposition~\ref{proposition: optimality} indicates that achieving asymptotic optimality requires keeping the steady-state expectations close to an optimal solution $y^{*,(n)}$ of~\eqref{eq:opt} for all sufficiently large $n$. 
However, designing schemes which can keep steady-state expectations close to target values is, in general, difficult unless the target values satisfy some structural properties.
In the following theorem, we investigate such structural properties by deriving closed form expressions for the optimal solutions of~\eqref{eq:opt}. The proof of the theorem is given in the appendix. 
It is important to emphasize here that the concavity and monotonicity of the speed-up function allow us to obtain such closed form solutions.

\begin{thm} 
\label{thm:optimization}
Assume that $\lambda^{(n)} \in (0,1]$ and the speed-up function of jobs satisfies~\eqref{prop:increasing} and~\eqref{prop:concavity}.
One of the following must hold.
\begin{enumerate}[(i)]
    \item If $\lambda^{(n)} \leq \frac{s_{d^{(n)}}}{d^{(n)}}$, then the optimal solution is unique and is given by $y^{*,(n)}=\left(0,0,\ldots,0,\frac{\lambda^{(n)}}{s_{d^{(n)}}}\right)$.
    
    \item If $\lambda^{(n)}=s_i/i$ for some $i \in [d^{(n)}]$, then the optimal solution $y^{*,(n)}=(y^{*,(n)}_i, i\in [d^{(n)}])$ of~\eqref{eq:opt} is unique and satisfies $y_{i_1}^{*,(n)}=\lambda^{(n)}/s_{i_1}$ and $y_j^{*,(n)}=0$ for all $j\neq i_1$, where $i_1=\max\{i\in [d^{(n)}]: \lambda^{(n)}=s_i/i\}$.
    
    \item If $\lambda^{(n)} \in \left(\frac{s_{i+1}}{i+1},\frac{s_i}{i}\right)$ for some $i\in [d^{(n)}-1]$ satisfying $\frac{s_{i+1}}{i+1}<\frac{s_i}{i}$, then an optimal solution $y^{*,(n)}=(y^{*,(n)}_i, i\in [d^{(n)}])$ of~\eqref{eq:opt} can be obtained by setting
    \begin{align}
        y_i^{*,(n)}&=\frac{\frac{1}{i}\brac{\lambda^{(n)}-\frac{s_{i+1}}{i+1}}}{\frac{s_i}{i}-\frac{s_{i+1}}{i+1}},\label{eq:first_comp}\\
        y_{i+1}^{*,(n)}&=\frac{\frac{1}{i+1}\brac{\frac{s_{i}}{i}-\lambda^{(n)}}}{\frac{s_i}{i}-\frac{s_{i+1}}{i+1}},\label{eq:second_comp}\\
        y_j^{*,(n)}&=0, \forall j \notin \{i,i+1\}\label{eq:other_comp}.
    \end{align}
    Furthermore, the above solution is unique if $s_i-s_{i-1} > s_{i+1} - s_{i} > s_{i+2}-s_{i+1}$.
\end{enumerate} 
\end{thm}
%

The above theorem shows how the structure of the optimal solution depends on the arrival rate $\lambda^{(n)}$ and the speed-up function $s$. 
In particular, it establishes that an optimal solution having at most two non-zero components always exists. 
This property, which is a consequence of the concavity and monotonicity of the speed-up function, implies that any scheme which aims to achieve asymptotic optimality must exhibit {\em state space collapse (SSC)} wherein the steady-state expected number of jobs receiving different numbers of servers lies in a subspace of dimension at most {\em two} within the original state-space of dimension $d^{(n)}$.
Henceforth, by $y^{*,(n)}$ we denote an optimal solution with at most two non-zero components and let $I^{*,(n)}=\{i: y^{*,{(n)}}_i > 0\}$ denote the indices of the non-zero components. 
Furthermore, we let $i_1^{(n)}$ and $i_2^{(n)}$ denote the minimum and maximum elements of $I^{*,(n)}$, respectively. From part (iii) of the theorem above, it follows that  $i_2^{(n)}\leq i_1^{(n)}+1$. Hence, an optimal scheme should drive the system to a steady state where each job receives either $i_1^{(n)}$ or $i_2^{(n)}$ servers since the steady-state expected number of jobs receiving $i \in [d^{(n)}]$ servers under the optimal solution is zero if $i\notin I^{*,{(n)}}$.

\begin{remark}
\label{rem:linear}
    Theorem~\ref{thm:optimization} implies that if the speed-up function is linear, i.e., if $s_i=i$ for all $i\in [d^{(n)}]$ and $\lambda^{(n)}\in (0,1]$, then $\abs{I^{*,(n)}}=1$, and $i_1^{(n)}=i_2^{(n)}=d^{(n)}$. The condition $\abs{I^{*,(n)}}=1$ also holds for sub-linear speed-up function as long as $\lambda ^{(n)} \leq s_{d^{(n)}}/d^{(n)}$ or $\lambda^{(n)} = s_i/i$ for some $i \in [d^{(n)}]$. In all other cases, we have $\abs{I^{*,(n)}}=2$. 
\end{remark}

\section{An Optimal Allocation Scheme}
\label{sec: main results}

The theorem in the previous section provides structural properties that the steady-state expectations should satisfy under an optimal scheme. In this section, we use these structural properties to construct a simple scheme under which the steady-state expectations converge to the optimal solution $y^{*,(n)}$ of~\eqref{eq:opt} as $n \to \infty$. 

To define the scheme, we first introduce the following notations. 
For each $i \in [d^{(n)}]$, define $p_i^{*,(n)}=s_iy_i^{*,(n)}/\lambda^{(n)}$. From the equality constraint of~\eqref{eq:opt}, it follows that $\sum_{i \in [d^{(n)}]}p_i^{*,(n)}=1$. Hence, $p^{*,(n)}=(p_i^{*,(n)}, i \in [d^{(n)}])$ is a valid probability vector.


{\bf The $\texttt{greedy}(p^{*,(n)})$ scheme}: If $j \in [0,n]$ servers are found available upon arrival of a job, then with probability $p_i^{*,(n)}, i \in [d^{(n)}]$, we allocate $\min(i,j)$ servers to the incoming job. We refer to this scheme as $\texttt{greedy}(p^{*,(n)})$ scheme as it allocates each job as many servers as possible up to $i\in [d^{(n)}]$ servers with probability $p_i^{*,(n)}$. Hence, in the $\texttt{greedy}(p^{*,(n)})$ scheme, the probability that an incoming job actually receives $i \in [d^{(n)}]$ servers when the system is in state $x$ is given by 
\begin{equation} \label{eq: assign_prob_sub_k=n}
   A_i^{(n)}(x) = \ind \brac{nq_0(x) \geq i} p_i^{*,(n)}+ \ind \brac{nq_0(x) = i}\sum_{j >i} p_j^{*,(n)}. 
\end{equation}
We define $A_0^{(n)}(x)$ to be the blocking probability in state $x$. Clearly, $\sum_{j=0}^{d^{(n)}} A_j^{(n)}(x) = 1$. 
We note that implementing the $\texttt{greedy}(p^{*,(n)})$ scheme requires the knowledge of the arrival rate and the speed-up function of jobs to compute the probability vector $p^{*,(n)}$. We assume that this knowledge is available. In practice, the arrival rate and the speed-up function can be easily learned from past data~\cite{autopilot_google,mao2016resource}.

{\bf The \texttt{greedy} scheme}: 
In addition to the above scheme, we introduce the $\texttt{greedy}$ scheme which always tries to assign as many servers as possible (up to a maximum of $d^{(n)}$) to each job.
It is worth noting that implementing this scheme does not require computing $p^{*,(n)}$. Hence, this scheme does not require knowledge of the arrival rate or the speed-up function of jobs. The $\texttt{greedy}(p^{*,(n)})$ scheme reduces to the \texttt{greedy} scheme when the probability vector $p^{*,(n)}$ is such that $p^{*,(n)}_d=1$ and $p^{*,(n)}_i=0$ for all $i < d^{(n)}$. According to the optimal solution given in Theorem~\ref{thm:optimization}, this occurs when the speed-up function is linear or the normalized arrival rate of jobs is smaller than $s_{d^{(n)}}/d^{(n)}$.

We now state our main result for the~$\texttt{greedy}(p^{*,(n)})$ scheme.
The results are divided into two cases: $|I^{*,(n)}|=1$ and $|I^{*,(n)}|=2$ (see Remark~\ref{rem:linear} for when these cases arise). 



\begin{thm} 
\label{thm: main thm}

Assume that $\lambda^{(n)}=1-\beta n^{-\alpha} \in (0,1)$ for $\alpha,\beta >0$. Furthermore, assume that $d^{(n)}=o(n)$. Then, under the $\texttt{greedy}(p^{*,(n)})$ scheme the following hold. 
 \begin{description}
     
 \item{i)} If $|I^{*,(n)}|=2$ and $\alpha \in (0,1)$, then  $$\mathbb{E}[\vert\vert x^{(n)}-y^{*,(n)}\vert\vert]=O\left(\frac{1}{{n}^{\min(1/4,(1-\alpha)/2)}}\right).$$ 
 Furthermore, both $P_b^{(n)}$ and $\abs{\mathbb{E}[D^{(n)}] -D^{*,{(n)}}}$ are  of the same order as $\mathbb{E}[\vert\vert x^{(n)}-y^{*,(n)}\vert\vert]$.

\item{i)} If $|I^{*,(n)}|=1$, then $\mathbb{E}[\vert\vert x^{(n)}-y^{*,(n)}\vert\vert]=O(1/\sqrt{n})$. Furthermore, we have
    \begin{align}
        P_b^{(n)}  \leq O\left(\frac{1}{{\sqrt{n}}}\right),
        \text{\quad and \quad } \abs{\mathbb{E}[D^{(n)}] -D^{*,{(n)}}}= O\left(\frac{1}{{n}}\right).
    \end{align}
\end{description}
Hence, the $\texttt{greedy}(p^{*,(n)})$ scheme is asymptotically optimal.
\end{thm}
The above theorem has several key implications. First, it indicates that as the system size $n$ tends to infinity, the steady-state expectations under the $\texttt{greedy}(p^{*,(n)})$ scheme converge to the optimal solution $y^{*,(n)}$, meeting the optimality criterion stated in Proposition~\ref{proposition: optimality}. 
Additionally, it offers non-asymptotic bounds, dependent on system size, for both the blocking probability and the mean execution time of jobs. These bounds characterize the rate at which convergence to optimality takes place under the proposed scheme. Notably, our bounds for $|I^{*,(n)}|=1$ are independent of the traffic parameters $\alpha$ and $\beta$ and holds for all $\alpha > 0$. We also note that $d^{(n)}=o(n)$ is sufficient to guarantee the optimality of the proposed scheme. Hence, even when the jobs are massively parallelizable with $d^{(n)} \to \infty$, the $\texttt{greedy}(p^{*,(n)})$ scheme is still asymptotically optimal as long as $d^{(n)}$ grows sufficiently slowly with $n$.

As a final remark, we note that the above theorem implies that the $\texttt{greedy}$ scheme is optimal when the speed-up function is linear or the arrival rate of jobs is sufficiently small since in both these cases the $\texttt{greedy}$ scheme is equivalent to the $\texttt{greedy}(p^{*,(n)})$ scheme. However, the optimality of the $\texttt{greedy}$ scheme no longer holds when
the speed-up function is sub-linear and the arrival rate is larger than $s_{d^{(n)}}/d^{(n)}$. In the latter case, 
the optimal number of servers to allocate to each job is given by the $\texttt{greedy}(p^{*,(n)})$ scheme with the probability vector $p^{*,(n)}$ computed using~\eqref{eq:first_comp} and~\eqref{eq:second_comp}.
\section{Proof of Theorem~\ref{thm: main thm}}
\label{section: proofs}

In this section, we prove Theorem~\ref{thm: main thm} using Stein's method. 
Since we consider a finite system of size $n$ throughout this section, we drop the superscript $\vphantom{\cdot}^{(n)}$ from our notations.


The core of our proof using Stein’s approach consists of comparing the dynamics of the system under the $\texttt{greedy}(p^{*})$ scheme to that of a simpler dynamical system. In this case, the simpler dynamical system is a {\em deterministic fluid limit} exhibiting the desired limiting behavior of the system 
which is defined through the following system of ODEs.
\begin{align} \label{eq: ode_sub}
    \dot{x}_i  =  \lambda p_i^* -s_i x_i=s_i y_i^*-s_i x_i, \quad i \in [d].
\end{align}
The above equations intuitively describe the evolution of a system with an infinite number of servers, where each arrival is allocated $i \in [d]$ servers with probability $p_i^* = s_i y_i^*/\lambda$, and jobs receiving $i$ servers depart at rate $s_i x_i$. Moreover, the trajectory of the fluid system converges to the desired optimal solution $y^*$ of~\eqref{eq:opt} as $t \to \infty$ starting from any initial state, capturing the desired behavior of our system under the $\texttt{greedy}(p^{*})$ scheme.
Our proof consists of showing that the steady-state of the original system under this scheme remains close to that of the fluid limit for large $n$.

To compare the steady-states of the two systems, we compare the generators of the original Markov process $x(\cdot)$ and the generator of the fluid limit~\eqref{eq: ode_sub} acting on a suitably chosen Lyapunov function. 
Let $G$ denote the generator of the Markov process $x(\cdot)$. In the steady-state, the expectation of the drift of any suitable function $V$ under $G$ is zero, i.e., $\expect{GV(x)} =0$.
Consider $L$ as the generator of the system of ODEs~\eqref{eq: ode_sub}. We then have $\expect{GV(x) - LV(x)} = \expect{-LV(x)}$. We choose the function $V(x)$ such that $-LV(x)$ corresponds to the mean squared distance between $x_i$ and $y^*_i$ for $i \in I^*$. Therefore, to analyze the deviation of the system state $x$ from its optimal value $y^*$, it is sufficient to bound the distance between the two generators acting on $V(x)$. The formal statement of this result is provided in Lemma~\ref{lemma: generator coupling_sub}.
\begin{lem} \label{lemma: generator coupling_sub}
For any positive constants $c_i>0, i \in I^*$, we have
\begin{equation}
\label{eq: first bound for general policy_sub}
\expect{\sum_{i \in I^*} c_i s_i \brac{x_i - y_i^*}^2} =  \frac{\lambda}{n} \sum_ {i\in I^*} c_i \expect{A_i(x)}
+
\sum_{i \in I^*} \expect{c_i \brac{x_i - y_i^*}\brac{\lambda A_i(x) - s_i y_i^*}}.
\end{equation}
\end{lem}
\begin{proof}
Let $G$ be the generator of the Markov process $x(\cdot)$ and $L$ be the generator of the system of ODEs given by~\eqref{eq: ode_sub}. We choose the function $V(x)$ in such a way that $-LV(x) = \sum_{i \in I^*} c_i s_i \brac{x_i - y_i^*}^2$, where $c_i$ is a strictly positive constant for every $i \in I^*$. 
Since $LV(x)=\sum_i \frac{\partial V}{\partial x_i} \dot x_i$, it is easy to see that the following choice of $V$ satisfies the above property.
\begin{equation}
    V(x)=\sum_{i \in I^*}\frac{c_i}{2}\brac{x_i - y_i^*}^2.
\end{equation}
Consequently, using $\mathbb{E}[GV(x)]=0$ we have
\begin{equation}
    \expect{GV(x) - LV(x)} = \expect{-LV(x)}= \expect{\sum_{i \in I^*} c_i s_i \brac{x_i - y_i^*}^2}.
\end{equation}
By definition of the generator $G$ of the Markov process $x(\cdot)$ we have
\begin{equation}\label{eq: drift G_sub}
GV(x)=\sum_{i \in I^*} n\lambda A_i(x)\brac{V\brac{x+\frac{1}{n}e_i}-V\brac{x}}
+\sum_{i \in I^*}ns_ix_i\brac{V\brac{x-\frac{1}{n}e_i}-V\brac{x}},
\end{equation}
where $e_i$ denotes the $d$-dimensional unit vector with a value of one at the $i^{th}$ position. Using Taylor series expansion of $V$, 
we have
\begin{align}
\expect{GV(x)-LV(x)}
&=\sum_{i \in I^*}\expect{n\lambda A_i(x)\brac{\frac{1}{n}\frac{\partial V}{\partial x_i}(x)+\frac{1}{2n^2}\frac{\partial^2 V}{\partial x_i^2}(\xi)}}\nonumber
\\&+\sum_{i \in I^*} \expect{ns_i x_i\brac{-\frac{1}{n}\frac{\partial V}{\partial x_i}(x)+\frac{1}{2n^2}\frac{\partial^2 V}{\partial x_i^2}(\theta)}-\frac{\partial V}{\partial x_i}(x)\dot{x}_i},
\end{align}
where $\xi$ and $\theta$ are $d$-dimensional vectors. 
Simplifying the RHS of the above and using the fact that $\frac{\partial^2 V}{\partial x_i^2}(y) = c_i$ for any vector $y$, we get
\begin{equation}
\mathbb{E}[GV(x)-LV(x)]=
\sum_{i \in I^*}\expect{\brac{\lambda A_i(x)-s_ix_i- \dot{x}_i}\frac{\partial V}{\partial x_i}(x)}
+\frac{1}{2n}\sum_{ i \in I^*} c_i\expect{\lambda A_i(x)+s_ix_i}.
\end{equation}
Finally, we arrive at the desired result by replacing $\dot x_i$ with the RHS of~\eqref{eq: ode_sub}, noting that $\frac{\partial V}{\partial x_i}(x) = c_i \brac{x_i - y_i^*}$, and using the rate conservation principle for jobs occupying $i$ servers which gives the equality $\expect{s_ix_i} = \lambda \expect{A_i(x)}$. 
\end{proof}
Hence, to show $\expect{\sum_{i \in I^*} c_i s_i \brac{x_i - y_i^*}^2} \to 0$, it is sufficient to bound the second term appearing on the RHS of~\eqref{eq: first bound for general policy_sub} for a suitable choice of the constants $c_i, i\in I^*$. In the rest of the proof, we choose $c_i=1$ for each $i\in I^*$ in~\eqref{eq: first bound for general policy_sub} and show that for $d=o(n)$ the following result holds.
\begin{proposition} \label{prop: full access}
Let $d = o(n)$. Then, for large enough $n$, we have 
\begin{equation}
\expect{\sum_{i \in I^*} s_i \brac{x_i - y_i^*}^2} \leq  \frac{1+2d}{n}
+ \ind\brac{\abs{I^*}=2}\brac{\frac{2d}{n}\brac{\frac{1}{\lambda}+\frac{1}{\frac{s_{i_1}}{i_1}-\frac{s_{i_2}}{i_2}}}+\frac{4d}{n\lambda\delta}+\frac{2\delta}{\frac{s_{i_1}}{i_1}-\frac{s_{i_2}}{i_2}}},
\end{equation}
where $\delta$ is any positive constant in the range $(0, 1-\lambda)$.
\end{proposition}

To prove the proposition above we make use of the following three key lemmas whose proofs are given in the Appendix.

\begin{lem} \label{lemma: SSC_two dim}
Under the equilibrium measure of the system, we have 
\begin{align}
    \mathbb{P} \brac{\sum_{i < i_1} x_i > 1/n} = 0,\label{eq: sum_less than i_1} \text{ and } \mathbb{P}\brac{\sum_{i>i_2} x_i  = 0} =1.
\end{align}
\end{lem}

\begin{lem} \label{lemma: second Lyapunov function,sublinear}
  For any $\delta \in \brac{0, 1- \lambda}$, define the Lyapunov function $V_2(x) = \ind \brac{r > \lambda+\delta}\sum_{i \in [d]} x_i$.
  Then, for any $\kappa >0$, we have 
  \begin{equation}
  \expect{V_2(x)} \leq \kappa+\frac{2}{ n\delta}, \text{ and } \mathbbm{P}\brac{r > \lambda + \delta} \leq \frac{s_d}{\lambda}\brac{\kappa+\frac{2}{ n\delta}},   
  \end{equation}
  as long as $d = o(n)$ and $n$ is sufficiently large. 
  \end{lem}

\begin{lem} \label{lemma: two comp_third bound}
For any constant $\delta \in (0, 1 - \lambda)$, the following holds for the case $\abs{I^*}=2$. 
\begin{equation}
 \expect{\ind\brac{q_1= 1- \frac{i_1}{n}, r \leq \lambda + \delta} \brac{(x_{i_1} - y_{i_1}^*)+(y_{i_2}^* - x_{i_2})}} 
         \leq \frac{1}{\frac{s_{i_1}}{i_1}-\frac{s_{i_2}}{i_2}} \brac{\frac{1}{n~i_2}(s_{i_1}+s_{i_2})+ \delta\brac{\frac{1}{i_1}+\frac{1}{i_2}}}.
\end{equation}
\end{lem}

\begin{figure}
\begin{center}
\includegraphics[width=0.5\linewidth]{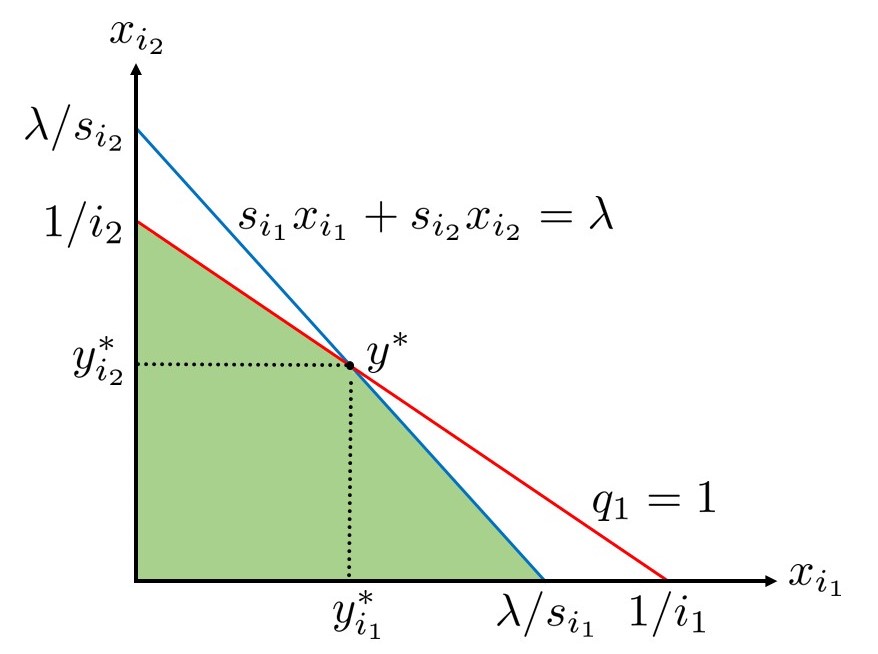}
\caption{State space of the system collapsing to two dimensions.} 
\label{fig: SSC}
\end{center}
\end{figure}

The three lemmas above characterize the region where the system operates with high probability in the steady-state. The first lemma indicates that for large system sizes the system is essentially two dimensional with only non-zero components given by $x_{i_1}$ and $x_{i_2}$. The second lemma shows that with high probability system operates in a region defined by the inequality $r\leq \lambda+\delta$ for small values of $\delta \in (0,1-\lambda)$ and large $n$. The last lemma shows that when $q_1=1-i_1/n$ in addition to $r\leq \lambda+\delta$, the system must operate near a region defined by the inequalities $x_{i_1}\leq y_{i_1}^*$
and $x_{i_2}\geq y_{i_2}^*$. Thus, when $q_1\approx 1$ and $r\leq \lambda+\delta$, the region of operation of the system is the intersection of the green region and the red line in Fig~\ref{fig: SSC}.

\subsection{Proof of Proposition~\ref{prop: full access}}

We first prove the following bound
on the second term on the RHS of~\eqref{eq: first bound for general policy_sub}
after choosing $c_i=1$ for all $i\in I^*$.
\begin{equation}\label{eq: three terms to bound}
   \sum_{i \in I^*} \expect{\brac{x_i - y_i^*} \brac{\lambda A_i(x) - s_i y_i^*}} 
   \leq 
   \frac{2d}{n} 
       +\brac{\frac{2d}{n\lambda}\brac{1+\frac{2}{\delta}} 
       +\frac{2}{\frac{s_{i_1}}{i_1}-\frac{s_{i_2}}{i_2}} \brac{\frac{d}{n}+ \delta}}\ind \brac{|I^*|=2},
  \end{equation}
where $\delta \in (0, 1- \lambda)$ is a constant. To prove the above bound, we consider the two cases of $|I^*|=1$ and $|I^*|=2$ separately. For $|I^*| = 1$, recall from Theorem~\ref{thm:optimization} that $y_{i_1}^* = \lambda / s_{i_1}$. Hence, using~\eqref{eq: assign_prob_sub_k=n} and $p_{i_1}^* = s_{i_1} y_{i_1}^* /\lambda$ we obtain
    \begin{align*} 
        \sum_{i \in I^*} \expect{ \brac{x_i - y_i^*} \brac{\lambda A_i(x) - s_i y_i^*}}  
        &=\lambda\expect{\brac{ y_{i_1}^*- x_{i_1} } \ind \brac{q_1 > 1-\frac{i_1}{n}}}\\
        &\leq \lambda \expect{\brac{ y_{i_1}^*- x_{i_1} } \ind \brac{q_1 > 1-\frac{i_1}{n}, y_{i_1}^*\geq x_{i_1}}}\\
        &\leq \lambda \expect{\brac{ i_1 y_{i_1}^*- i_1 x_{i_1} } \ind \brac{q_1 > 1-\frac{i_1}{n}, y_{i_1}^*\geq x_{i_1}}}\\
        &\leq \lambda \expect{\brac{ 1- q_1 + \sum_{i < i_1} ix_i} \ind \brac{q_1 > 1-\frac{i_1}{n}, y_{i_1}^*\geq x_{i_1}}} \leq \frac{2 \lambda i_1}{n}\leq \frac{2 d}{n},
    \end{align*}
%
%
where the first inequality on the last line follows from the facts that $i_1 y_{i_1}^* \leq \sum_{i \in [d]} i y_i^* \leq 1$ and, for $\abs{I^*}=1$, $i_1 x_{i_1}=q_1 - \sum_{ i < i_1} ix_i$, and the second inequality on the last line follows from
Lemma~\ref{lemma: SSC_two dim} which implies $\sum_{i < i_1} ix_i < i_1/n$.
    
Using a similar line of arguments for the case where $|I^*|=2$, we obtain the following after some simplification.
\begin{align}    
   \sum_{i \in I^*} \expect{\brac{x_i - y_i^*} \brac{\lambda A_i(x) - s_i y_i^*}} 
    & \leq \expect{ \ind\brac{q_1 > 1- \frac{i_1}{n}, y_{i}^*\geq x_i, \forall i \in I^*} \brac{1-q_1+\sum_{i < i_1} i x_i}} \label{eq: first term_sub}
    \\&+ \expect{\ind \brac{q_1 =1 - \frac{i_1}{n}}\brac{(x_{i_1} - y_{i_1}^*)+(y_{i_2}^* - x_{i_2})}}\label{eq: third term_sub}.
\end{align}
Expression~\eqref{eq: first term_sub} can be bounded by $2d/n$ as before.
To bound \eqref{eq: third term_sub}, we decompose it as follows.
\begin{align} 
    \eqref{eq: third term_sub}
     & = \mathbb{E}\biggl[\ind\brac{q_1 = 1- \frac{i_1}{n}, r > \lambda + \delta} 
    \brac{(x_{i_1} - y_{i_1}^*)+(y_{i_2}^* - x_{i_2})}\biggr]\label{eq: third term_case1_sub}
    \\& +\mathbb{E}\biggl[\ind\brac{q_1 = 1- \frac{i_1}{n}, r \leq \lambda + \delta} 
   \brac{(x_{i_1} - y_{i_1}^*)+(y_{i_2}^* - x_{i_2})}\biggr],\label{eq: bound on third term_case2_sub}
\end{align}
where $\delta \in (0 , 1- \lambda)$ is a constant. Now, we have
\begin{equation}
   \eqref{eq: third term_case1_sub}
    \leq 2 \mathbb{P}\brac{ r > \lambda + \delta} \leq \frac{2s_d}{n\lambda}\brac{1+\frac{2}{\delta}}\leq \frac{2d}{n\lambda}\brac{1+\frac{2}{\delta}},
\end{equation}
where the second inequality follows from Lemma~\ref{lemma: second Lyapunov function,sublinear} by choosing $\kappa=1/n$ and the last inequality follows from $s_d \leq d$. Finally, \eqref{eq: bound on third term_case2_sub} can be bounded using Lemma~\ref{lemma: two comp_third bound} and noting that 
$s_{i_1}+s_{i_2}\leq 2d$ to complete the proof of~\eqref{eq: three terms to bound}.

Now to complete the proof of Proposition~\ref{prop: full access}, we note from Lemma~\ref{lemma: generator coupling_sub} that for $c_{i_1}=c_{i_2}=1$, the first term on the RHS of~\eqref{eq: first bound for general policy_sub} is bounded by $1/n$.



\subsection{Proof of Theorem~\ref{thm: main thm}} 

We first note the following.
\begin{align}
\expect{\vert \vert x - y^*\vert\vert} &= \expect{\sum_{ i\not\in I^*} \abs{x_i}} + \expect{\sum_{ i\in I^*} \abs{x_i - y^*_i}}\nonumber\\
& \leq \expect{\sum_{ i\not\in I^*} x_i} + \sqrt{2\expect{\sum_{ i\in I^*} \brac{x_i - y^*_i}^2}}\nonumber\\
&\leq \expect{\sum_{ i\not\in I^*} x_i} + \sqrt{2\expect{\sum_{ i\in I^*} s_i \brac{x_i - y^*_i}^2}}\nonumber\\
&\leq \frac{1}{n}+\sqrt{\frac{2+4d}{n}
+\ind\brac{\abs{I^*}=2}2\brac{\frac{2d}{n}\brac{\frac{1}{\lambda}+\frac{1}{\frac{s_{i_1}}{i_1}-\frac{s_{i_2}}{i_2}}}+\frac{4d}{n\lambda\delta}+\frac{2\delta}{\frac{s_{i_1}}{i_1}-\frac{s_{i_2}}{i_2}}}},\nonumber
\end{align}
where the first line follows from the definition of the set $I^*$, the second line follows from Jensen's inequality and the fact that $(a+b)^2\leq 2(a^2+b^2)$ for all $a,b\in \mb{R}$, the third line follows from the fact that $s_i \geq 1$ for all $i \in [d]$, and the last line follows from Lemma~\ref{lemma: SSC_two dim} and Proposition~\ref{prop: full access}. We recall that $\delta \in (0,1-\lambda)$.
Hence, when $\lambda$ varies according to $\lambda=1-\beta n^{-\alpha}$, we must choose $\delta \in (0, \beta n^{-\alpha})$. When $\alpha \in (0,1/2)$, choosing $\delta=\beta/2\sqrt{n}$ gives the lowest possible order of $O(1/n^{1/4})$ for the second term in the last inequality above. However, when $\alpha > 1/2$, the same choice of $\delta$ does not work since the condition $\delta \in (0, \beta n^{-\alpha})$ is violated. Hence, we choose $\delta=\beta/2n^{\alpha}$ which gives the order $O(1/n^{(1-\alpha)/2})$ for the second term on the last line. Thus, combining the above bounds we have
\begin{equation}
\expect{\vert \vert x - y^*\vert\vert}=O\brac{\frac{1}{\sqrt{n}}}+\ind\brac{\abs{I^*}=2}O\brac{\frac{1}{n^{\min(1/4,(1-\alpha)/2)}}}.
\end{equation}

To derive a bound on the blocking probability, we use the rate conservation law which yields 
\begin{equation}
    \lambda\brac{1- P_b} = \expect{r} = \sum_{i \in [d]} s_i\expect{x_i}.
\end{equation}
Given the non-negativity of $x_i$ for $i\in [d]$, it follows that $\sum_{i \in [d]} s_i x_i \geq \sum_{i \in I^*} s_i x_i$. Thus,
\begin{equation}
    \lambda P_b \leq \expect{\brac{\lambda - \sum_{i \in I^*}s_{i} x_{i}}} = \sum_{i \in I^*}s_{i} \expect{ y_i^* -x_{i}} \leq d \expect{\norm{x-y^*}},
\end{equation}
where the equality follows from the fact that $\sum_{ i\in I^*} s_i y_i^* = \lambda$ and the last inequality follows from the fact that $s_i\leq d$ for all $i \in [d]$. Hence, the bound on the blocking probability has the same order as that of $\expect{\norm{x-y^*}}$.

Similarly, for the mean response time of accepted jobs, we use the Little's law to have
\begin{align}
    \lambda \brac{1-P_b} \expect{D}  = \sum_{i \in [d]} \expect{x_i}.
\end{align}
Furthermore, since $\lambda D^* = \sum_{i \in [d]} y_i^*$, we have 
\begin{align}
    \lambda  (\expect{D}-D^*)=  &\sum_{i\in [d]}\frac{1}{1-P_b}\expect{x_i-y_i^*}+\frac{P_b}{1-P_b}\sum_{i \in [d]}y_i^*.\nonumber
\end{align}
Taking absolute value on both sides of the above equation and using the fact that $\sum_{i \in [d]}y_i^* \leq \sum_{i \in [d]}i y_i^* \leq 1$, we have
\begin{equation}
\lambda  \abs{\expect{D}-D^*}\leq  \frac{1}{1-P_b}\expect{\norm{x-y^*}}+\frac{P_b}{1-P_b}.
\end{equation}
From the above it follows that $\abs{\expect{D}-D^*}$ is of the same order as $P_b$ and $\expect{\norm{x-y^*}}$.


For the case $|I^*|=1$, we can derive a stronger bound on $\abs{\expect{D}-D^*}$ as follows. From Little's law, we have
\begin{equation}
    \lambda(1-P_b)\expect{D} = \sum_{i \in [d]} \expect{x_i} \leq \frac{1}{n} + \expect{x_{i_1}},
\end{equation}
where the inequality follows from Lemma~\ref{lemma: SSC_two dim}. Noting that $D^* = 1/s_{i_1}$ in this case, we have
\begin{equation*}
    \lambda(1-P_b)\expect{D} \leq \frac{1}{n} + D^*\expect{s_{i_1}x_{i_1}} \leq  \frac{1}{n} + D^*\expect{r} =  \frac{1}{n} + D^*\lambda(1-P_b),
\end{equation*}
where the second inequality follows from the non-negativity of $x_i$ for $i \in [d]$ and the equality follows from the rate conservation law. Therefore, we have $\expect{D} \leq \frac{1}{n\lambda(1-P_b)} + D^*$. This inequality combined with the fact that $\expect{D} \geq D^*$ gives the desired result.

\section{Heterogeneous Workloads}
\label{sec:het}

Thus far we have only considered systems where all jobs have the same speed-up function. However, our results can be easily generalized to systems with multiple job classes with different speed-up functions. To see this, let us consider a system with $l$ job classes indexed by $j \in [l]$. As before we drop the superscript $\vphantom{\cdot}^{(n)}$ from our notations for brevity. Let class $j$ jobs be parallelizable up to $d_j$ servers with a speed-up function $s_j=(s_{i,j}, i \in [d_j])$. We assume that the speed-up function for each class satisfies properties \eqref{prop:increasing} and \eqref{prop:concavity}. Then, using the same line of arguments as in Section~\ref{sec:lower}, we have that the optimal scheme which tries to minimize the mean execution time of jobs under the constraint of zero blocking must solve the following optimization problem

\begin{equation}
        \begin{aligned}
        & \underset{y=(y_{ij})} {\text{minimize}}
        & & \frac{1}{\lambda}\sum_{i,j} y_{i,j}\\
        & \text{subject to}
        & & \sum_{i \in [d_j]} s_{i,j} y_{i,j} = \frac{\lambda_j}{\mu_j}, \forall j \in [l],\\
        & & &\sum_{i,j} i y_{i,j} \leq 1,\\
        & & &  y_{i,j} \geq 0, \quad \forall j \in [l], i \in [d_j]
        \end{aligned}
        \tag{Q}
    \label{eq:opt_het}
\end{equation}
where $y_{i,j}$ denotes the (scaled) expected number of class $j$ jobs occupying $i$ servers in the steady-state; $\lambda_j$ and $\mu_j$ respectively denote the arrival rate and mean inherent job size of class $j$ and $\lambda = \sum_j \lambda_j$. Defining $\rho_j=\lambda_j/\mu_j$, it is easy to see that the above optimization problem is equivalent to finding the fractions $b_j \in [0,1]$, $j \in [l]$,  of servers to reserve for use of class $j$ jobs such that they solve the following optimization problem

\begin{equation}
        \begin{aligned}
        & \underset{b=(b_{j})} {\text{minimize}}
        & & \sum_{j} f_j(b_j)\\
        & \text{subject to}
        & & \sum_{j} b_j \leq 1,\\
        & & &  b_{j} \geq \rho_j, \quad \forall j \in [l],
        \end{aligned}
        \tag{Q1}
    \label{eq:opt_het_outer}
\end{equation}
where $f_j:[0,1]\to \mathbb{R}$ for each class $j \in [l]$ is defined as the optimal objective function value of the following single-class problem for class $j$

\begin{equation}
        \begin{aligned}
        & f_j(b):==\underset{y=(y_{i,j}, i\in [d_j])}{\min}
        & & \frac{1}{\lambda}\sum_{i\in [d_j]} y_{i,j}\\
        & \text{subject to}
        & & \sum_{i \in [d_j]} s_{i,j} y_{i,j} = \rho_j, \\
        & & &\sum_{i\in [d_j]} i y_{i,j} \leq b,\\
        & & &  y_{i,j} \geq 0, \quad \forall i \in [d^{(n)}],
        \end{aligned}
        \tag{Q2}
    \label{eq:opt_het_inner}
\end{equation}
Note that the single class optimization problem~\eqref{eq:opt_het_inner}  is equivalent to~\eqref{eq:opt} (after re-defining the decision variables as $y_{i,j}/b$). Hence, the function $f_j$ can be computed in closed form using the results of Theorem~\ref{thm:optimization} and plugging this expression in~\eqref{eq:opt_het_outer}, we can compute the optimal fractions $b_j^*$, $j \in [l]$ of servers to reserve for each class. Once the optimal proportion $b_j^*$ of servers to reserve for each class $j$ has been found in this way, the optimal solution $y_{i,j}^*$ of the original problem~\eqref{eq:opt_het} can be found by solving~\eqref{eq:opt_het_inner} with $b$ replaced by $b_j^*$. It is easy to see that when $\sum_j \rho_j < 1$, there exists a feasible solution to~\eqref{eq:opt_het}. Furthermore, using the $\texttt{greedy}(p_j^*)$ scheme with $p_{i,j}^*=s_{i,j}y_{i,j}^*/\rho_j$ within each class $j$ will lead to asymptotic optimality.  The proof of asymptotic optimality remains the same as the single-class case because,  within each class, the structure of the optimal solution (in particular the SSC property which makes the proof of Theorem~\ref{thm: main thm} work) remains the same as given in Theorem~\ref{thm:optimization} and the classes can be treated independently once we have decided to reserve $b_j^*$ fraction of servers for class $j$ jobs for each $j \in [l]$.
\section{Numerical Results} \label{section: numerical results}
In this section, we present simulation results that validate our analytical findings. 
The simulations are conducted $100$ times, each simulating the arrival of the first $5$ million jobs, with a maximum degree of parallelism set at $d^{(n)}=5$. Two types of speed-up functions are considered: linear (where $s_1=1$, $s_2=2$, $s_3=3$, $s_4=4$, $s_5=5$) and sub-linear (where $s_1=1$, $s_2=1.8$, $s_3=2.5$, $s_4=3$, $s_5=3.4$). Various traffic regimes are simulated, including the mean-field regime with $\alpha=0, \beta = 0.2$, Halfin-Whitt regime with $\alpha=1/2, \beta=0.1$, and super-Halfin-Whitt regime with $\alpha=2/3, \beta=0.1$. Furthermore, 
two allocation schemes of $\texttt{greedy}(p^{*,(n)})$ and $\texttt{greedy}$, defined in Section~\ref{sec: main results} are considered. 

Figure~\ref{fig: blocking,probabilistic} depicts the blocking probability of the system. It is evident that with the $\texttt{greedy}(p^{*,(n)})$ scheme, the blocking probability converges to zero as the system size $n$ increases. However, when employing the $\texttt{greedy}$ scheme with the sub-linear speed-up function, the system exhibits a non-zero blocking probability. This observation highlights the consequence of allocating more servers to each job which can result in a persistent non-zero blocking probability. 

Moreover, Figure~\ref{fig: response,probabilistic} presents the mean execution time of accepted jobs. Under the $\texttt{greedy}(p^{*,(n)})$ scheme, the mean execution time of accepted jobs converges to $ D^{*,(n)}= \frac{1}{\lambda^{(n)}}\sum_{i \in I^{*,(n)}} y_i^{*,(n)}$, where the optimal solution $y^{*,(n)}$ is given in Theorem~\ref{thm:optimization}. 
Specifically, when the speed-up function is linear, $y^{*,(n)} = \brac{0,0,0,0,\lambda^{(n)}/s_5}$, resulting in $D^{*,(n)} = 0.2$. For the sub-linear speed-up function, with $\alpha=0$ and $\beta = 0.2$, the optimal solution $y^{*,(n)}$ is given by $\brac{0, 0,0.2,0.1,0}$ and $D^{*,(n)}=0.375$. For all other cases when the speed-up function is sub-linear and $\alpha>0$, it holds that $y^{*,(n)} \to \brac{1, 0, 0, 0, 0}$, and therefore $D^{*,(n)} \to 1$. Furthermore, it is observed that when employing a sub-linear speed-up function and the $\texttt{greedy}$ scheme, the system achieves a lower average execution time compared to the $\texttt{greedy}(p^{*,(n)})$ scheme 
; however, at the expense of a non-zero blocking probability, indicating that the scheme is no longer asymptotically optimal. 

\begin{figure}
\subfigure[Blocking probability of the system]{\label{fig: blocking,probabilistic}\includegraphics[width=.5\linewidth]{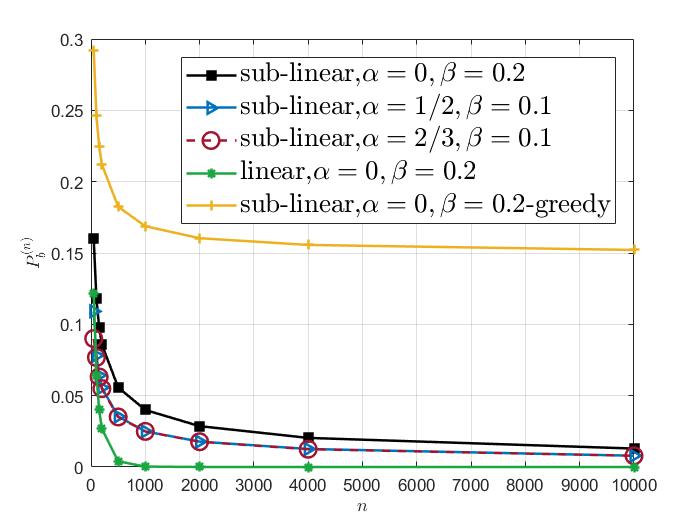}}\hfill
\subfigure[Mean response time of accepted jobs]{\label{fig: response,probabilistic}\includegraphics[width=.5\linewidth]{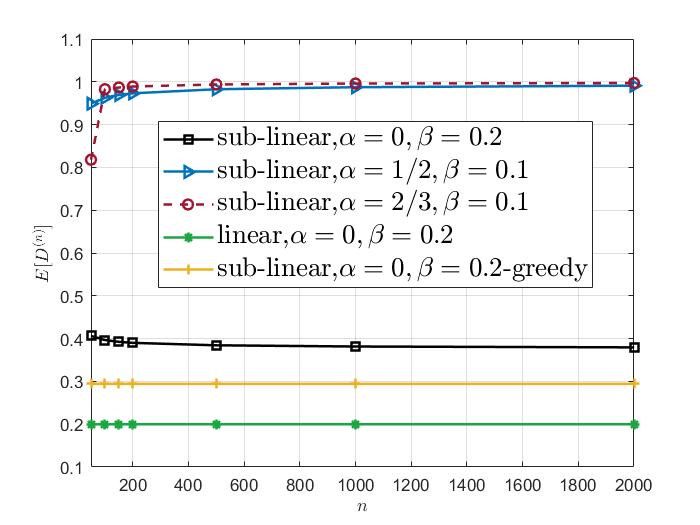}}
\caption{
system performance metrics for different system sizes $n$ under $\texttt{greedy}(p^{*,(n)})$ and $\texttt{greedy}$ schemes}
\label{fig: system performance,probabilistic}
\end{figure}

%

Finally, we study the insensitivity of the $\texttt{greedy}(p^{*,(n)})$ scheme to the exact distribution of job sizes in Table~\ref{tabel:general size}. We consider the same system parameters 
and study a system with $n=4000$ servers, considering exponential (Exp), deterministic (Det), Mixed-Erlang, and Pareto distributions, all with the same unit mean. The Mixed-Erlang distribution includes two exponential phases with probabilities $p_1=0.4$, and $p_2=0.6$. The CDF of the Pareto distribution is given by
$\mathbbm{P} \brac{Y \leq y} = \begin{cases}
     1- \frac{1}{(3y)^{3/2}},\quad &\text{ if } y \geq \frac{1}{3}\\
     0, \quad &\text{otherwise}
\end{cases}$. Results in Table~\ref{tabel:general size} show the system's performance is insensitive to the exact distribution of job lengths.

\renewcommand{\arraystretch}{0.7}
\begin{table*}[htbp]
\caption{Performance metrics of a system with $n=4000$ servers under different job size distributions}
\small
\makebox[\textwidth][c]{\begin{tabular}{c|m{3.5em}m{3.5em}m{3.5em}m{3.5em}|m{3.5em}m{3.5em}m{3.5em}m{3.5em}}
\toprule
 &\multicolumn{4}{c|}{Mean Execution Time $\expect{D^{(n)}}$} &  \multicolumn{4}{c}{Blocking Probability $P_b^{(n)}$}\\
\midrule
Inherent Size Distribution & Exp & Det & Mixed Erlang& Pareto&Exp & Det & Mixed Erlang& Pareto\\
\midrule
linear speed-up, $\alpha=0, \beta = 0.2$ &0.2000&0.2000&0.2000&0.1973&0& 0& 0&0 \\
 linear speed-up, $\alpha=1/2, \beta = 0.1$ &0.2000&0.2000&0.2000&0.1970&0.0267&0.0268&0.0267&0.0209 \\
 linear speed-up, $\alpha=2/3, \beta = 0.1$ &0.2000&0.2000&0.2000&0.1971&0.0274&0.0274&0.0274&0.0219  \\
\hline
sub-linear speed-up, $\alpha=0, \beta = 0.2$ &0.3782&0.3782&0.3782&0.3708&0.0204&0.0202&0.0203&0.0149 \\
sub-linear speed-up, $\alpha=1/2, \beta = 0.1$ &0.9930&0.9937&0.9933& 0.9621&0.0126&0.0126&0.0126& 0.0041\\
sub-linear speed-up, $\alpha=2/3, \beta = 0.1$ &0.9976&0.9984&0.9979&0.9669&0.0125&0.0125&0.0125&0.0041 \\
\bottomrule
\end{tabular}
}
\label{tabel:general size}
\end{table*}

\section{Conclusion} \label{section: conclusion}

The present paper opens up many directions for future research. One immediate problem to address is to analytically establish the insensitivity of the system operating under the $\texttt{greedy}(p^{*,(n)})$ scheme. This will require studying the system under more general distributions. Another interesting direction to pursue is to consider multi-phase jobs with each phase having a different speed-up function. Obtaining optimal schemes for such a system in the stochastic setting is a challenging open problem. Furthermore, our scheme requires the knowledge of the arrival rate and the speed-up functions of jobs to make optimal server allocations. Deriving schemes which can automatically learn these parameters will be an interesting problem to consider.
%
\bibliographystyle{unsrt}
\bibliography{MobiRef_trunc3.bib}
\appendix

\section{Proof of Proposition~\ref{proposition: optimality}}
\label{proof:optimality}
First, we note that for $\lambda^{(n)} \in [0,1]$, the $d^{(n)}$-dimensional vector $(\lambda^{(n)},0,\ldots,0)$ is a feasible solution of~\eqref{eq:opt}. Furthermore, by comparing the first two constraints of~\eqref{eq:opt}, it is clear that for there to be a feasible solution to~\eqref{eq:opt}, we must have $\lambda^{(n)} \leq 1$ since by~\eqref{eq:concave} we have $s_i \leq i$ for all $i \in [d^{(n)}]$.

Subtracting~\eqref{eq:rate_consevation} from the first constraint of problem~\eqref{eq:opt}
we obtain $$r(y^{*,(n)})-\expect{r^{(n)}}=\sum_{i\in [d^{(n)}]}s_i \expect{y^{*,(n)}_i-x_i^{(n)}}=\lambda^{(n)} P_b^{(n)}.$$
Hence, taking the absolute values of both sides, using the triangle inequality, and the fact that $s_i \leq d^{(n)}$ for all $i \in [d^{(n)}]$, we obtain $\lambda^{(n)} P_b^{(n)} \leq d^{(n)} \expect{\norm{x^{(n)}-y^{*,(n)}}}$.
Thus, if $\expect{\norm{x^{(n)}-y^{*,(n)}}} \to 0$ for a scheme, then $P_b^{(n)}\to 0$ for the same scheme. Now, we note that $\lambda^{(n)}\expect{D^{(n)}}=\frac{\sum_{i \in [d^{(n)}]}\expect{x_i^{(n)}}}{1-P_b^{(n)}}$ and ${\lambda^{(n)}}{D^{*,(n)}}={\sum_{i \in [d^{(n)}]}{y_i^{*,(n)}}}$. 
Hence, we have
$\abs{\lambda^{(n)}(\expect{D^{(n)}}-D^{*,(n)})}\leq \frac{1}{1-P_b^{(n)}}\expect{\norm{x^{(n)}-y^{*,(n)}}}+\frac{P_b^{(n)}}{1-P_b^{(n)}}.$ The desired result follows from the above inequality since  $P_b^{(n)}\to 0$ and $\expect{\norm{x^{(n)}-y^{*,(n)}}} \to 0$.

\section{Proof of Theorem~\ref{thm:optimization}}
\label{proof: optimization proposition}

To simplify notations, we drop the superscript $\vphantom{\cdot}^n$ from all the notations used in the proof. The Lagrangian function for~\eqref{eq:opt} is given by

\begin{equation*}
    \mathcal{L}\brac{y,\nu, \theta_0, \theta_1, \ldots, \theta_d} = \frac{1}{\lambda}\sum_{i \in [d]} y_i + \nu \brac{\sum_{i \in [d]} s_i y_i - \lambda}+ \theta_0 \brac{\sum_{i \in [d]} i y_i -1} - \sum_{i \in [d]} \theta_i y_i,    
\end{equation*}
where $\nu \in \mb R$, $\theta_0 \geq 0$, and $\theta_i \geq 0$, $i\in [d]$ denote the Lagrange multipliers associated with the equality constraint, the first inequality constraint and the non-negativity constraints of~\eqref{eq:opt}, respectively. Slater's condition for strong duality holds since $y = \brac{\lambda, 0, \ldots, 0}$ is a feasible solution for all $\lambda \in (0,1]$. Consequently, any primal optimal solution $y = \brac{y_1, \ldots, y_d}$ and dual optimal solution $\brac{\nu, \theta_0, \theta_1,\ldots,\theta_d}$ must satisfy Karush-Kuhn-Tucker (KKT) conditions given below.

\begin{align}
    &\frac{\partial{\mathcal{L}}}{\partial{y_i}} = \frac{1}{\lambda} + \nu s_i + \theta_0 i - \theta_i =0, \forall i \in [d],  \label{eq: zero derivative}
    \\ &\theta_0 \geq 0, \quad \theta_i \geq 0, \forall i \in [d], \quad   \label{eq: dual feas}
    \\& \theta_0 \brac{\sum_{i \in [d]} i y_i -1}=0, \quad \theta_i y_i=0, \forall i \in[d], \quad \label{eq: slackness}
    \\& \sum_{i \in [d]} s_i y_i  = \lambda, \quad \sum_{i \in [d]} i y_i \leq 1, \quad y_i \geq 0, \forall i \in [d]. \quad  \label{eq: primal feas}
\end{align}
From the primal feasibility constraint  $\sum_{i \in [d]} s_i y_i = \lambda >0$ (in \eqref{eq: primal feas}), it follows that $y_i>0$ for at least one $i \in [d]$. The complementary slackness condition $\theta_i y_i =0$ (in \eqref{eq: slackness}) requires  $\theta_i=0$ if $y_i >0$. Let $\theta_i =0$ (or, equivalently, $y_i > 0$) for $K \geq 1$ distinct indices of $i \in \{i_1,i_2,\ldots,i_K\} \subseteq [d]$ with $i_1 < i_2 < \ldots <i_K$, and $\theta_i >0$ for all other indices. Hence, by~\eqref{eq: zero derivative} we have

\begin{align} \label{eq: zero thetai}
    \frac{1}{\lambda}+ \nu s_{i_k} + \theta_0 i_k = \theta_{i_k} = 0, \quad \forall k \in [K].
\end{align}

We consider two cases: one where $\theta_i = 0$ for at least two distinct indices of $i$, i.e., $K \geq 2$, and the other where $\theta_i = 0$ for a single index $i$, i.e., $K=1$. 

\begin{enumerate}
   \item \textbf{The $K\geq 2$ case:} In this case, we must have $\nu\neq 0$ as $\nu=0$ implies by~\eqref{eq: zero thetai} that $\theta_0=-1/i_k \lambda$ for at least two distinct values of $i_k$ which is not possible. Hence, $\nu\neq 0$ and this by~\eqref{eq: zero thetai} implies
\begin{equation} \label{eq: theta0 / nu}
    -\frac{\theta_0}{\nu} = \frac{s_{i_{k_2}} - s_{i_{k_1}}}{i_{k_2} - i_{k_1}}, \quad \forall k_1 <k_2,~ k_1, k_2 \in [K].
\end{equation}
 We show that all indices $\{i_1,i_2,\ldots,i_K\}$ must be consecutive, i.e., $i_{k+1}=i_k+1$ for all $k \in [K-1]$. Assume it is not true. Therefore, $i_k + 1 < i_{k+1}$ for some $ k \in [K]$. Then, the interval $\brac{i_k, i_{k+1}}$ contains at least one index $i \in [d]$. Furthermore, by definition of the set of indices $\{i_1,i_2,\ldots,i_K\}$, for any $i \in \brac{i_k, i_{k+1}}$, we must have $\theta_i >0$. Consequently, using~\eqref{eq: zero derivative} for $i$ and $i_k$ we have

\begin{align}
   &\frac{1}{\lambda} + \nu s_i + \theta_0 i  = \theta_i >0, \quad \forall i \in \brac{i_k, i_{k+1}}.
\end{align}
Using the above and~\eqref{eq: zero thetai} we have
\begin{equation}
    -\frac{\theta_0}{\nu} > \frac{s_i - s_{i_k}}{i - i_k},\quad \forall i \in \brac{i_k, i_{k+1}}.
\end{equation}
Combined with~\eqref{eq: theta0 / nu} the above yields

\begin{equation}
    \frac{s_{i_{k+1}} - s_{i_k}}{i_{k+1} - i_k} >\frac{s_i - s_{i_k}}{i - i_k}, \quad \forall i \in \brac{i_k, i_{k+1}},
\end{equation}
which is equivalent to

\begin{equation}
    s_i < \brac{1- \frac{i - i_k}{i_{k+1}-i_k}} s_{i_k} + \frac{i-i_k}{i_{k+1} - i_k} s_{i_{k+1}}, 
\end{equation}
for all $i \in \brac{i_k, i_{k+1}}$. Since $\frac{i - i_k}{i_{k+1}-i_k} \in (0,1)$ for any $i \in \brac{i_k, i_{k+1}}$, the above inequality violates the concavity of the speed-up function. Therefore, we conclude that

\begin{equation} \label{eq: consecutive index}
    i_{k+1} = i_k +1, \quad \forall k \in [K-1].
\end{equation}
Applying the above in~\eqref{eq: theta0 / nu}, we conclude that 

\begin{equation} \label{eq: constant diff Delta}
    s_{i_{k+1}} - s_{i_k} = -\frac{\theta_0}{\nu}, \quad \forall k \in [K-1],
\end{equation}
This implies that $\theta_0 >0$ as otherwise \eqref{eq: constant diff Delta} leads to $s_{i_{k+1}}=s_{i_k}$ for all $k \in [K-1]$, contradicting~\eqref{eq:increasing}. 
For simplicity, we shall denote the ratio $-\theta_0/\nu$ by $\Delta$. Note that by~\eqref{eq: zero thetai} we have
 \begin{equation}
            s_{i_K} - \Delta i_K = s_{i_{K-1}} - \Delta i_{K-1} = \ldots = s_{i_1} - \Delta i_1.
            \label{eq:identity}
\end{equation}

   Since $\theta_0>0$, by the complementary slackness condition $\theta_0 \brac{\sum_{i \in [d]} i y_i -1 }=0$, we obtain $\sum_{i \in [d]} i y_i =1$. Further, since $y_i$ is non-zero only for $i \in \{ i_1, i_2, \ldots, i_K \}$, we have

   \begin{align}
       \sum_{k=1}^K s_{i_k} y_{i_k} &= \lambda,\label{eq:rate}
       \\\sum_{k=1}^K i_k y_{i_k} &= 1.\label{eq:capacity}
   \end{align}
   However, from~\eqref{eq: consecutive index} and \eqref{eq: constant diff Delta}, we have $i_k = i_1 + (k-1)$ and $s_{i_k} = s_{i_1} +  (k-1)\Delta$ for any $k \in [K]$. As a result, we obtain



    \begin{align}
        \brac{s_{i_1} - \Delta i_1} \sum_{k=1}^K y_{i_k} &= \lambda - \Delta,
        \label{eq: sum first K}\\ \sum_{k=1}^K (k-1) y_{i_k} &= 1- i_1 \sum_{k=1}^K y_{i_k}. \label{eq: sum first K scaled}
    \end{align}

    We note that $\Delta = \lambda$ is not possible since it implies by~\eqref{eq: sum first K} that $s_{i_1}-\Delta i_1 =0$. But, by~\eqref{eq:identity} this implies $\frac{s_{i_k}}{i_k} = \Delta = \lambda$, for all $k \in [K]$ substituting which in~\eqref{eq: zero thetai} yields
        $$\nu \lambda + \theta_0 = -\frac{1}{\lambda i_k}, \quad \forall k \in [K].$$
    However, the above cannot hold for more than one distinct values of $i_k$. Hence, $\Delta \neq \lambda$ which by~\eqref{eq: sum first K}-\eqref{eq: sum first K scaled} implies that

        \begin{align}
            \sum_{k=1}^K y_{i_k} &= \frac{\lambda - \Delta}{s_{i_1} - \Delta i_1}, \label{eq: first K sum subcase2}
            \\
            \sum_{k=2}^K (k-1) y_{i_k} &= 1- i_1\sum_{k=1}^K y_{i_k} = \frac{s_{i_1}-\lambda i_1}{s_{i_1}-\Delta i_1}. \label{eq: first K-1 sum subcase2}
        \end{align}

        We now claim that $s_{i_1}-\Delta i_{1} > 0$ and therefore by~\eqref{eq:identity} $s_{i_k}-\Delta i_k >0$ for all $k \in [K]$.
        Assume that the above is not true, i.e., $s_{i_k}-\Delta i_k < 0$ for all $k \in [K]$. Then, by~\eqref{eq: constant diff Delta} we obtain
        \begin{align}
            \frac{s_{i_k}}{i_k} < \Delta  = s_{i_{k+1}} - s_{i_k}, \quad \forall k \in [K-1].
            \label{eq:strict}
        \end{align}
        But since $i_{k+1}=i_k+1$ for all $k \in [K-1]$, the above implies $s_{i_{k}}/i_k < s_{i_{k+1}}/i_{k+1}$ which contradicts~\eqref{eq:concave}. Hence, We conclude that $s_{i_k}-\Delta i_k >0$ for all $k \in [K]$. This shows that the direction of inequality in~\eqref{eq:strict} should be reversed and therefore the following must hold.
        \begin{equation}
            \frac{s_{i_1}}{i_1} > \frac{s_{i_2}}{i_2} > \ldots > \frac{s_{i_K}}{i_K}.
        \end{equation}
        But the above implies that $$\sum_{k\in [K]} s_{i_k} y_{i_k}=\sum_{k\in [K]} (s_{i_k}/i_k) i_k y_{i_k} \in \brac{\frac{s_{i_K}}{i_K},\frac{s_{i_1}}{i_1}}$$ since $\sum_{k \in [K]} i_k y_{i_k}=1$ by~\eqref{eq:capacity}.
        Hence, by~\eqref{eq:rate} we must have
        $\lambda \in \brac{\frac{s_{i_K}}{i_K},\frac{s_{i_1}}{i_1}}$.
        Hence, to summarize, we have shown the following: for $K\geq 2$ if there exists $\{i_1,i_2,\ldots,i_K\}\subseteq[d]$ such that
        \begin{itemize}
            \item $i_{k+1}=i_k+1, \forall k\in [K-1]$,
            \item $s_{i_{k+1}}-s_{i_k} =\Delta $ for some  constant $\Delta >0$ and all $k \in [K-1]$,
            \item $s_{i_{k}}/i_k > s_{i_{k+1}}/i_{k+1}, \forall k \in [K-1]$,
            \item and $\lambda \in \brac{\frac{s_{i_K}}{i_K},\frac{s_{i_1}}{i_1}}$,
        \end{itemize}
        then any 
        $y\in \mathbb{R}^d_+$ which satisfies~\eqref{eq: first K sum subcase2},~\eqref{eq: first K-1 sum subcase2},  and $y_i=0$ for all $i\notin \{i_1,i_2,\ldots,i_K\}$ is a primal optimal solution of~\eqref{eq:opt}. In particular,
        \begin{align*}
            y_{i_k}&=\frac{\frac{1}{i_k}\brac{\lambda-\frac{s_{i_{k+1}}}{i_{k+1}}}}{\frac{s_{i_k}}{i_k}-\frac{s_{i_{k+1}}}{i_{k+1}}}, \\ 
            y_{i_{k+1}}&=\frac{\frac{1}{i_{k+1}}\brac{\frac{s_{{i_k}}}{i_k}-\lambda}}{\frac{s_{i_k}}{i_k}-\frac{s_{i_{k+1}}}{i_{k+1}}},\\
            y_i&=0, \forall i \notin \{i_{k},i_{k+1}\}.
        \end{align*}
        constitutes one such optimal solution. Furthermore, when $K=2$, the above is the unique optimal solution. This establishes the last part of Theorem~\ref{thm:optimization}.

    \item \textbf{The $K=1$ case:} Let $i_1 \in [d]$ be the only index for which $\theta_{i_1} =0$. Therefore, $\theta_i >0$ for all $i \neq i_1$ and the complementary slackness condition $\theta_i y_i=0$ in~\eqref{eq: slackness} implies $y_{i}=0$ for all $i \not= i_1$. As a result, from primal feasibility constraint $\sum_{i \in [d]} s_{i} y_i = \lambda$, we obtain
    \begin{align*}
        y_{i_1} = \frac{\lambda}{s_{i_1}},\quad y_{i} = 0,~ \forall i \not = i_1.
    \end{align*}

    Two sub-cases emerge based on whether $\lambda < s_{i_1} / i_1$ or $\lambda = s_{i_1} / i_1$.
    \begin{itemize}
        \item Sub-case $1$, $\lambda< s_{i_1} / i_1$: In this sub-case,  $\sum_{i \in [d]} i y_i  = i_1 y_{i_1}=\frac{i_1\lambda}{s_{i_1}}<1$ which implies $\theta_0 = 0$. Consequently, from~\eqref{eq: zero derivative} we obtain
    \begin{align*}
        &1/\lambda+ \nu s_{i_1} = \theta_{i_1}=0
        \\ &1/\lambda+ \nu s_i = \theta_i >0, \quad\forall i \not= i_1.
    \end{align*}

    \item Sub-case $2$, $\lambda =  s_{i_1} / i_1$: In this sub-case, the only non-zero component of $y$ is given by 
    $$
    y_{i_1}  =  \frac{\lambda}{s_{i_1}} = \frac{1}{i_1}.$$
    Hence, the objective function becomes $1/\lambda i_1$. Therefore, if there exists multiple indices $i \in [d]$ for which $\lambda=s_i/i$, then the objective function will be minimized only if we choose $i_1 = \max \{i: \lambda   = s_{i} / i \}$. 
     \end{itemize}
     The above two sub-cases together completes the proof of the first two parts of Theorem~\ref{thm:optimization}.
\end{enumerate}

\section{Proof of Lemma~\ref{lemma: SSC_two dim}}

To prove the lemma, it is sufficient to assume the system starts at a state where $\sum_{ i\not\in I^*}x_i=0$ since, due to ergodicity, starting from any other state the system will reach a state satisfying the above condition in a finite time with probability one. Now, if the system starts at a state satisfying $\sum_{ i\not\in I^*}x_i=0$, then the $\texttt{greedy}(p^*)$ scheme will keep allocating each incoming job either $i_1$ or $i_2$ free servers until the number of free servers in the system falls strictly below $i_1$. Until this point, the value of $x_{i}$ for $i \notin I^*$ will remain zero. Once the number of free servers in the system falls strictly below $i_1$, the next arrival increases $\sum_{i < i_1} x_i$ from zero to at most $1/n$. This job is referred to as the {\em tagged job}. If, upon arrival of the tagged job, $\sum_{i < i_1} x_i$ increases to $1/n$, then the system must have become fully occupied after the arrival of the tagged job. This is because of the fact that the $\texttt{greedy}(p^*)$ scheme allocates all available servers to an incoming job if the number of available servers goes below $i_1$. Hence, until the tagged job leaves the system, subsequent arrivals either find the system fully busy (and are therefore blocked) or find at least $i_1$ servers available (which occurs if a job occupying $i_1$ or $i_2$ servers departs before the arrival occurs and the tagged job leaves the system). In either case, the sum $\sum_{i < i_1} x_i$ remains constant at $1/n$ until the tagged job departs. Upon the departure of the tagged job, $\sum_{i < i_1} x_i$ returns to zero. From this point onward we can apply the same chain of arguments as above. This shows that $\sum_{i < i_1} x_i$ never exceeds $1/n$. Furthermore, at all times, there is no increase in the number of components $x_i$ for $i > i_2$ because under the $\texttt{greedy}(p^*)$ the probability of a job being allocated more than $i_2$ servers is zero.

\section{Proof of Lemma~\ref{lemma: second Lyapunov function,sublinear}}

Assume that the system is in a state where $V_2(x)  \geq \kappa$ for some $\kappa>0$. By definition of $V_2$, this can only happen when $r > \lambda+\delta$. An arrival can only increase the value of $r$ and a departure can decrease it by at most $s_d/n \leq d/n$. Hence, when $V_2(x)  \geq \kappa > 0$, $d=o(n)$, and $n$ is sufficiently large, the value of $r$ after an arrival or a departure  still remains above $\lambda+\delta$.
%
Hence, given $V_2(x)  \geq \kappa$, the drift of the function $V_2(x)$ under $G$ for sufficiently large $n$ satisfies
\begin{align}
    GV_2(x) = \lambda \sum_{i \in [d]} A_i(x)  - \sum_{i \in [d]} s_ix_i
     \leq  \lambda - r 
    < -\delta,
\end{align}
where the second inequality follows from the fact that $r > \lambda +\delta$ (since $V_2(x)  \geq \kappa$). 
We now apply Lemma~4 of~\cite{lei_ying_2020_Stein} 
to conclude that
\begin{align} \label{eq: expect_v2}
     \expect{V_2(x)} \leq  \kappa+\frac{2}{ n\delta}.
\end{align}
This completes the proof of the first part of the lemma. For the second part, note that
\begin{align*}
   \lambda \mathbbm{P}\brac{r > \lambda + \delta}  &=  \lambda \expect{\ind \brac{r > \lambda + \delta}}
  \leq \expect{\ind \brac{r > \lambda + \delta} r} 
   \nonumber\\& \leq s_d \expect{\ind \brac{r > \lambda + \delta} \sum_{i \in [d]}x_i}
   \leq  s_d\brac{\kappa+\frac{2}{ n\delta}},
\end{align*}
where the first inequality in the second line follows since $r = \sum_{i \in [d]} s_i x_i \leq s_d \sum_{i\in[d]} x_i$, and the last inequality follows from the definition of $V_2(x)$ and~\eqref{eq: expect_v2}. This completes the proof.

\section{Proof of Lemma~\ref{lemma: two comp_third bound}}

Given that $|I^*|=2$, $q_1=1-i_1/n$, and $r \leq \lambda +\delta$,  we have the following using the definitions of $q_1$ and $r$.
\begin{align}
    i_1 x_{i_1} + i_2 x_{i_2} +\sum_{i < i_1} ix_i&= 1 - \frac{i_1}{n},\label{eq:elim1}
    \\s_{i_1}x_{i_1} + s_{i_2}x_{i_2} +\sum_{i <i_1}s_ix_i& \leq \lambda + \delta.
    \label{eq:elim2}
\end{align}
By eliminating $x_{i_2}$ from the above we obtain
\begin{align}
    {i_1}x_{i_1} \brac{\frac{s_{i_1}}{i_1}-\frac{s_{i_2}}{i_2}} - (\lambda -\frac{s_{i_2}}{i_2}) &\leq \delta + \frac{i_1 s_{i_2}}{n i_2}+\sum_{i < i_1} \brac{\frac{s_{i_2}}{i_2}-\frac{s_{i}}{i}} ix_i\\
    &\leq \delta + \frac{i_1 s_{i_2}}{n i_2},
\end{align}
where the last line follows since $\frac{s_{i_2}}{i_2}\leq \frac{s_i}{i}$ for all $i <i_1$ by~\eqref{eq:concave}.
%
%
Now, using $y_{i_1}^* = \frac{\frac{1}{i_1} \brac{ \lambda- \frac{s_{i_2}}{i_2}}}{\frac{s_{i_1}}{i_1}-\frac{s_{i_2}}{i_2}}$ we obtain
\begin{equation}
(x_{i_1}-y_{i_1}^*)\leq \frac{\frac{\delta}{i_1} + \frac{s_{i_2}}{n i_2}}{\frac{s_{i_1}}{i_1}-\frac{s_{i_2}}{i_2}}.
\end{equation}
Similarly, eliminating $x_{i_1}$ from~\eqref{eq:elim1} and~\eqref{eq:elim2} we obtain
\begin{equation}
(y_{i_2}^*-x_{i_2})\leq \frac{\frac{\delta}{i_2} + \frac{s_{i_1}}{n i_2}}{\frac{s_{i_1}}{i_1}-\frac{s_{i_2}}{i_2}}.
\end{equation}
Combining the above two we obtain the result of the lemma.

\end{document}